\documentclass[english,aps,preprint,superscriptaddress]{revtex4}
\usepackage[]{fontenc}
\usepackage{amsmath}
\usepackage{graphicx}
\usepackage{bm}% bold math

\makeatletter
\@ifundefined{textcolor}{}
{%
 \definecolor{BLACK}{gray}{0}
 \definecolor{WHITE}{gray}{1}
 \definecolor{RED}{rgb}{1,0,0}
 \definecolor{GREEN}{rgb}{0,1,0}
 \definecolor{BLUE}{rgb}{0,0,1}
 \definecolor{CYAN}{cmyk}{1,0,0,0}
 \definecolor{MAGENTA}{cmyk}{0,1,0,0}
 \definecolor{YELLOW}{cmyk}{0,0,1,0}
 }

\makeatletter

\newcommand{\be}{\begin{equation}}\newcommand{\ee}{\end{equation}}\newcommand{\ba}{\begin{align}}\newcommand{\ea}{\end{align}}\def\bea{\begin{eqnarray}}\def\eea{\end{eqnarray}}

\usepackage{slashed}

\makeatother

\usepackage{babel}

\begin{document}

\title{TeV Scale Phenomenology of $e^+e^- \to\mu^+ \mu^-$ Scattering in the Noncommutative Standard Model with Hybrid Gauge Transformation }

\author{Weijian Wang}
\affiliation{Department of Physics, Zhejiang University, Hangzhou
310027, P.R. China}
\author{Jia-Hui Huang}
\affiliation{Center of Mathematical Science, Zhejiang University,
Hangzhou 310027, P.R.China}
\author{Zheng-Mao Sheng}\thanks{corresponding author's Email:zmsheng@zju.edu.cn}
\affiliation{Department of Physics, Zhejiang University, Hangzhou
310027, P.R. China}

\begin{abstract}
The hybrid gauge transformation and its nontrivial
phenomenological implications are investigated using the
noncommutative gauge theory with the Seiberg-Witten map expanded
scenario. Particularly, the $e^+e^- \to\mu^+ \mu^-$ process is
studied with a generalized noncommutative standard model (NCSM)
including massive neutrinos and neutrino-photon interaction. In
this model, the hybrid gauge transformation in the lepton sector
is naturally introduced through the requirement of gauge
invariance of the seesaw neutrino mass term. It is shown that in
the NCSM without hybrid gauge transformation  the noncommutative
correction to the scattering amplitude of the $e^+e^- \to\mu^+
\mu^-$ process appears only  as a phase factor, predicting no new
physical deviation in the cross section. However, when the hybrid
feature is considered, the noncommutative effect appears in the
single channel process. The cross section and angular distribution
are analyzed in the laboratory frame including Earth's rotation.
It is proposed that pair production of muons in the upcoming TeV
International Linear Collider (ILC) can provide an ideal
opportunity for exploring not only the NC space-time, but also the
mathematical structure of the corresponding gauge theory.

\end{abstract}
\maketitle

\section{Introduction}

Although we are still far from a complete theory unifying quantum
mechanics and general relativity, the noncommutative (NC)
space-time is a common feature appearing in many existing theories
of quantum gravity. The concept of noncommutative space-time was
first introduced in Snyder's pioneer work \cite{snyder1947}.
Interest in noncommutative space-time has been revived in the
recent decades due to its connection to string theory
\cite{seiberg1999, Connes1998} (for a review, see
\cite{Douglas2001}). It is generally believed that the stringy
effect can only be observed at the Planck scale $M_{P}$. However,
given the scenario suggested by the extra-dimension theories
\cite{extra} that the large hierarchy between Planck scale $M_{P}$
and the weak scale $M_{W}$ can be strongly reduced, one can expect
to see the NC effect at TeV scale, which is detectable in the LHC
and other planned colliders. A popular noncommutative model is
that the NC space-time is characterized by the coordinate operator
satisfying
\begin{equation}
[\hat{x}_{\mu},\hat{x}_{\nu}]=i\theta_{\mu\nu}=\frac{ic_{\mu\nu}}{\Lambda^2_{NC}},
\label{homer}
\end{equation}
where the matrix matrix  $\theta_{\mu\nu}$ is constant,
antisymmetric and real, in units of $(\mathrm{mass})^{-2}$. The
elements of the dimensionless constant matrix $c_{\mu\nu}$ are
assumed to be of order unity, and $\Lambda_{NC}$ represents the NC
scale. One can decompose the NC parameters $\theta_{\mu\nu}$ into
two classes: electric-like component
$\theta_{E}=(\theta_{01},\theta_{02},\theta_{03})$ associated with
time-space noncommutativity and magnetic-like component
$\theta_{B}=(\theta_{23},\theta_{31},\theta_{12})$ associated with
space-space noncommutativity. Through the Weyl correspondence, the
quantum field theory in NC space-time can be equivalent to that in
commutative space-time with the ordinary product of field
variables replaced by the Weyl-Moyal star product \cite{moyal}
\begin{equation}
\phi_{1}*\phi_{2}(x)=\exp(\frac{i}{2}\theta^{\mu\nu}\partial_{\mu}^{x}\partial_{\nu}^{y})\phi_{1}(x)\phi_{2}(y)|_{y\rightarrow
x}.\label{eq:moyalproduct}\end{equation} Using this method, the
QED in noncommutative space-time (NCQED) has been constructed and
extensively studied by many authors (for a review, see
\cite{Szabo2003}). However, to build a NC extension of the
standard model (NCSM), one encounters some obstructions, such as
charge quantization \cite{Haya} and the no-go theorem
\cite{Chaichian mms}. Up to now, a minimal version of the
noncommutative standard model (NCSM) has been proposed in Ref.
\cite{Calmet1}, in which the consistency problem mentioned above
is overcome when one generalizes the SU(3)*SU(2)*SU(1) Lie algebra
gauge theory to the enveloping algebra value using the
Seiberg-Witten map (SWM) method \cite{seiberg1999}. The SWM means
that there is a map between the noncommutative fields and their
classical counterparts as a power series expanded in $\theta$
\begin{eqnarray}
\hat{\psi}(x,\theta)&=&\psi(x)+\theta\psi^{(1)}+\theta^2\psi^{(2)}+...  \label{bart}                      \\
\hat{A}_{\mu}(x,\theta)&=&A_{\mu}(x)+\theta
A^{(1)}_{\mu}+\theta^2A^{(2)}_{\mu}+... \label{lisa}
\end{eqnarray}
where $\hat{\psi}$ and $\hat{A}_{\mu}$ denote the fields in NC
space-time. The NCSM predicts not only NC correction of particle
vertex but also new interactions beyond the standard model in
ordinary space-time, e.g. $Z-\gamma-\gamma$ and $Z-g-g$ vertices
\cite{behr03}. The rich phenomenological implications have led to
intense studies of various high energy processes \cite{behr03,
Das08, Sheng05,Pra2010, W2011}.

In the construction of NCSM, the so-called hybrid gauge
transformation and hybrid SWM of Higgs fields are adopted to
ensure covariant Yukawa terms \cite{Calmet1}. In this scenario,
the Higgs fields feel a "left" charge and a "right" charge in NC
gauge theory. Although only applied to the Higgs sector in Ref.
\cite{Calmet1}, the method can in principle be extended to all
other fields. One of the extensions has resulted in  notable new
physics predicted by NCQED: the tree-level interaction between
neutrino and photon \cite{Sch04}. In NCQED, the interaction
between fermion and photon are of three types:
$e\hat{A}_{\mu}*\Psi$, $e\Psi *\hat{A}_{\mu}$ and
$e(\hat{A}_{\mu}*\Psi-\Psi *\hat{A}_{\mu})$. The first two
interaction are charge conjugated of each other. One can also
consider it as the ambiguity in the ordering of Weyl-Moyal
product. The third coupling is particularly interesting. In this
case, the neutral particle transforms under NC U(1) gauge field
from the left and right sides in a similar way as in the adjoint
representation in the ordinary non-Abelian gauge theory. The
covariant derivative is
\begin{equation}
\hat{D}_{\mu}\hat{\Psi}=\partial_{\mu}\hat{\Psi}-i[\hat{A_{\mu}},
\hat{\Psi}]_{*}.
\end{equation}
Then the action is invariant when  $\Psi$ encounters the hybrid
gauge transformation
\begin{equation}
\hat{\Psi}^{\prime}=U*\hat{\Psi}*U^{-1}
\end{equation}
where $U=(e^{i\Lambda})_{*}$. The phenomenology of photon-neutrino
interaction has been extensively explored \cite{Sch04, MME2006}.
It is well known that one can not construct interactions such as
$2e\hat{A}_{\mu}*\Psi-e\Psi *\hat{A}_{\mu}$ in the
 context of Lie algebra because the covariant derivatives can only be applied
to the fermion fields of charged 0, $\pm1$ \cite{Haya}. However,
as  mentioned above, this restriction can be broken by extending
the group structure from Lie algebra to the enveloping one with
the help of SWM, as discussed in Sec. 2. We shall see below that
this will lead to interesting phenomenological implication.

On the other hand, the NCSM in Ref. \cite{Calmet1} is constructed
without including the neutrino mass. However, neutrino oscillation
experiments have provided  convincing evidence of massive neutrinos
and lepton favor mixing \cite{neu2002}. Thus it is natural to
question if the massive neutrinos and its direct interaction with
photon as mentioned above can accommodate each other in the
framework of NCSM. The issue has been studied in Ref. \cite{MM2008}.
It is found that such an extension does not work for massive Dirac
neutrino, while  massive Majorana neutrinos are still consistent
with the gauge symmetry. This means we have to accept the
photon-neutrino interaction at the cost of ruling out the popular
seesaw mechanisms \cite{neu1,neu2} that successfully generates Dirac
neutrinos with small mass scale in the standard model and Majorana
neutrinos with the GUT mass scale. In a recent work \cite{RH2011},
the authors showed that the difficulty presented in Ref.
\cite{MM2008} can be overcome by appropriately generalizing the NC
gauge transformation and SWM to a hybrid formation. In this sense,
we can construct a generalized NCSM including the seesaw model and
neutrino-photon interaction. The authors in Ref. \cite{RH2011}
derived the Feynman rules of photon-neutrino interaction and Z
boson-neutrino interaction in the NCSM incorporated with type I
seesaw mechanism. As an phenomenological application, the $Z$ boson
decays were studied in a very recent work\cite{Zdecay}.

It is interesting to investigate if  generalization of the NCSM
has any nontrivial effect on the phenomenology. A first choice is
to explore the distinct neutrino-photon interaction which has been
studied by many authors \cite{Sch04, MME2006}. In this paper,
however, we focus our attention on a simple high energy process
$e^+e^- \to\mu^+ \mu^-$. The processes has been studied in Ref.
\cite{Pra2010} using the NC corrected Feynman rules up to
$\theta^{2}$ order. It has been shown that after considering all
orders of the Seiberg-Witten map, the NC correction to the $e^+e^-
\to\mu^+ \mu^-$  appears only as phase factors, leaving no net
noncommutative effect \cite{W2011}. In the generalized mNCSM,
things are different. As we shall see later, the covariant
derivatives of leptons require modifications to guarantee the
gauge invariance of the Dirac-type mass term due to the presence
of photon-neutrino interaction. We shall see that the
modifications will have impact on the lowest order gauge coupling
of charged leptons and eventually lead to a nontrivial NC
correction for the scattering cross section. In Sec. 2, we first
introduce the hybrid gauge transformation by considering the
simplest case: the NCQED with $U(1)$ Abelian group. Then, we
briefly review the NCSM incorporated with massive neutrino and
neutrino-photon interaction given in Ref. \cite{RH2011}. The
relevant Feynman rules involving all orders of the NC parameter
$\theta$ are derived. In Sec. 3, we give the scattering amplitude
of $e^+e^- \to\mu^+ \mu^-$ in the laboratory frame where the earth
rotational effect is considered. Numerical analyses of the total
cross section and angular distribution are presented in Sec.4. We
summarize our results in Sec.5.

\section{HYBRID GAUGE TRANSFORMATION IN NC GAUGE THEORIES}

\subsection{Hybrid Gauge Transformation in Noncommutative Abelian Theory }
For simplicity, we start by investigating the Abelian NC $U(1)$
gauge theory. In this case,  the NC Lagrangian for a fermion
$\hat{\Psi}$ is
\begin{equation}
\hat{S}_{NC}=\int d^4x[i\bar{\hat{\Psi}}\gamma^{\mu}\hat{D}_{\mu
}\hat{\Psi}-m\bar{\hat{\Psi}}\hat{\Psi}],
\label{matter}\end{equation} where
$\hat{D}_{\mu}\hat{\Psi}=\partial_{\mu}\hat{\Psi}-ie\hat{A_{\mu}}*
\hat{\Psi}$. The action is invariant under the gauge
transformation
\begin{equation}
\hat{\Psi}^{\prime}(x,\theta)=U*\hat{\Psi}(x,\theta),
\label{trans}\end{equation}
\begin{equation}
\hat{A}_{\mu}^{\prime}(x,\theta)=U*\hat{A}_{\mu}(x,\theta)*U^{-1}+\frac{i}{e}U*U^{-1},
\end{equation}
where $U=(e^{i\Lambda})_{*}$. From the view point of gauge
invariance, there is no $\textsl{priori}$ requirement that we must
take Eq.\eqref{trans} as the only possible representation. In the
enveloping algebra formulation, the NC gauge theory works well for
arbitrary charges. With the help of SWM, one can extend
Eq.\eqref{trans} to the so-called hybrid formation in which the
spinor $\hat{\Psi}$ proceeds under both "left" and "right"
transformation:
\begin{equation}
\hat{\Psi}^{\prime}(x,\theta)=U_{L}*\hat{\Psi}(x,\theta)*U_{R}^{-1}
\end{equation}
with $U_{L}=(e^{i\Lambda})_{*}$ and
$U_{R}=(e^{i\Lambda^{\prime}})_{*}$. Then the corresponding
covariant derivative is
\begin{equation}
\hat{D}_{\mu}\hat{\Psi}=\partial_{\mu}\hat{\Psi}-i(e+e^{\prime})\hat{A}_{\mu
L}* \hat{\Psi}+ie^{\prime}\hat{\Psi}*\hat{A}_{\mu R},
\label{cor}\end{equation} where we define the "left (right)" NC
gauge fields $\hat{A}_{\mu L}$($\hat{A}_{\mu R}$) transforming as
\begin{equation} \hat{A}_{\mu
L}^{\prime}(x,\theta)=U_{L}*\hat{A}_{\mu
L}(x,\theta)*U_{L}^{-1}+\frac{i}{e+e^{\prime}}U_{L}*U_{L}^{-1},
\end{equation}
\begin{equation}
\hat{A}_{\mu R}^{\prime}(x,\theta)=U_{R}*\hat{A}_{\mu
R}(x,\theta)*U_{R}^{-1}+\frac{i}{e^{\prime}}U_{R}*U_{R}^{-1}.
\end{equation}
One can think of $\hat{\Psi}$ as having a "left" charge
$e+e^{\prime}$ and a "right" charge $e^{\prime}$. However,
$\hat{A}_{\mu L}$ and $\hat{A}_{\mu R}$ are  gauge fields not for
different particles but for the different NC representations of
SWM of the ordinary gauge potential $A_{\mu}$. Up to the zeroth
order of $\theta$, their expressions of SWM  are the same:
$\hat{A}_{\mu L}\simeq A_{\mu}\simeq\hat{A}_{\mu R}$. When the
limit $\theta\rightarrow0$ is taken, the NC covariant derivative
Eq. \eqref{cor} reduces to the ordinary one with the right
electro-charge in commutative space-time. The hybrid feature
presented here is derived from the degrees of freedom of NC gauge
theory. The exact value of $e^{\prime}$ can not be constrained by
the NC gauge invariance itself. In the existing literature, the
$e^{\prime}$ is set to  zero and the electron field $\hat{\Psi}$
only transforms as simplest representation Eq. \eqref{trans}. We
believe that existence of more subtle representation is possible,
and explore the phenomenological implication of it. The argument
used in Abelian case can easily be extended to more realistic
models. In the subsection B, we use a generalized NCSM as proposed
in Ref. \cite{RH2011}, where the massive neutrinos and the
photon-neutrino interaction is incorporated.

\subsection{Hybrid Gauge Transformation in Noncommutative Standard Model}
In this subsection, we briefly review the NCSM generalized by the
seesaw mechanism and photon-neutrino interaction. Following the
Ref. \cite{RH2011}, the action of the generalized NCSM is

\begin{equation}
\hat{S}_{GNCSM}=\hat{S}_{gauge}+\hat{S}_{quark}+\hat{S}_{lepton}+\hat{S}_{Higgs}
+\hat{S}_{Yukawa},
\end{equation}
where the gauge  and quark sectors are the same as that in the
NCSM of Ref. \cite{Calmet1}. However, we will see that the lepton,
Higgs, and Yukawa sectors are modified to incorporate the seesaw
mechanism and neutrino-photon interaction. In this paper, we only
take the simplest type-I seesaw model into account, but the
conclusion should be qualitatively applicable to other types. For
our purpose, the Higgs and Yukawa sectors of the leptons are
\begin{equation}\begin{split}
\hat{S}_{Higgs}=&\int d^4x
[(\hat{D}_{\mu}\hat{\Phi}_{d})^{\dagger}*(\hat{D}^\mu\hat{\Phi}_{d})-\mu^2\hat{\Phi}_{d}^{\dagger}*
\hat{\Phi}_{d}-\lambda\hat{\Phi}_{d}^\dagger*\hat{\Phi}_{d}*\hat{\Phi}_{d}^\dagger*\Phi_{d}]\\
&+\int
d^4x[(\hat{D}_{\mu}\hat{\Phi}_{s})^{\dagger}*(\hat{D}^\mu\hat{\Phi}_{s})-\mu^2\hat{\Phi}_{s}^{\dagger}*
\hat{\Phi}_{s}-\lambda\hat{\Phi}_{s}^\dagger*\hat{\Phi}_{s}*\hat{\Phi}_{s}^\dagger*\Phi_{s}],
\end{split}\end{equation}
\begin{equation}\begin{split}
\hat{S}_{Yukawa}&=\hat{S}_{Dirac}+\hat{S}_{Majorana}\\
&=-\int d^4x \sum_{i,j=1}^3
[y_{ij}(\bar{\hat{\Psi}}_{L}^{i}*\hat{\Phi}_{d}*\hat{l}_R^{j})
 +
 y^{\dagger}_{ij}(\bar{\hat{l}}_R^{i}*{\hat{\Phi}}_{d}^\dagger*\hat{\Psi}_{L}^{j})\\
 &+y^{\prime}_{ij}(\bar{\hat{\Psi}}_L^{i}*\hat{\Phi}_{d}^{c}*\hat{\nu}_R^{j})+
 y^{\prime\dagger}_{ij}(\bar\hat{\nu}_{R}^{i}*\hat{\Phi}_{d}^{c\dagger}*\hat{\Psi}_L^{j})]\\
 &-\frac{i}{2}\int d^4x \sum_{i,j=1}^3
 [t_{ij}(\hat{\nu}_{R}^{i
 T}*\hat{\Phi}_{s}^{c}*\sigma_{2}\hat{\nu}^{j}_{R})
 -t_{ij}^{\dagger}(\hat{\nu}^{iT}_{R}*\hat{\Phi}_{s}^{c\dagger}*\sigma_{2}\hat{\nu}^{j}_{R})],
\label{yukawa}
\end{split}\end{equation}
where we have denoted the noncommutative left handed doublet of
leptons, right handed singlet lepton, right handed neutrino,
doublet Higgs boson and singlet Higgs fields  respectively as
\begin{equation}
\hat{\Psi}_{L}=\begin{pmatrix}{\hat{\nu}_{L}}\\{\hat{l}_{L}}\end{pmatrix},\quad\quad\hat{l}_{R},\quad\quad\hat{\nu}_{R},\quad\quad
\hat{\Phi}_{d},\quad\quad\hat{\Phi}_{s},
\end{equation}
respectively, and $i$, $j$ are the generation indices, and
$y_{ij}$, $y^{\prime}_{ij}$ and $t_{ij}$ are Yukawa coupling
constants. It is noted that in the generalized NCSM, we need not
only the doublet Higgs fields but also a singlet Higgs fields to
ensure the Majorana mass term of the right handed neutrino.

In ordinary space-time, the neutral-hyper-charged $\hat{\nu}_{R}$
singlet does not directly couple to any gauge field. However, in
NC space-time, it can be coupled to the $U_{*}(1)$ hyper gauge
field $\hat{B}_{\mu}$ through a star commutator
\begin{equation}
D_{\mu}\hat{\nu}_{R}=\partial_{\mu}\hat{\nu}_{R}-i\kappa
g_{Y}[\hat{B}_{\mu},\hat{\nu}_{R}]_{*},
\end{equation}
and $\hat{\nu}_{R}$ transforms under noncommutative $U_{*}(1)$ gauge
group as
\begin{equation}
\delta_{\hat{\Lambda}}\hat{\nu}_{R}=i\kappa
g_{Y}\hat{\Lambda}*\hat{\nu}_{R}-i\kappa
g_{Y}\hat{\nu}_{R}*\hat{\Lambda},
\end{equation}
where $\hat{\Lambda}$ is the gauge parameter and $\kappa$ is an
unknown multiple or fractional number of the coupling constant. To
ensure  gauge invariance of the Yukawa sector Eq. \eqref{yukawa},
one can see that the transformation rules of the left-handed
lepton doublet, right-handed charged lepton singlet, Higgs
doublet, and Higgs singlet are respectively modified to
\begin{equation}\begin{split}
&\delta_{\hat{\Lambda}}\begin{pmatrix}{\hat{\nu}_{L}}\\{\hat{l}_{L}}\end{pmatrix}
=ig_{Y}[(-\frac{1}{2}+\kappa)\hat{\Lambda}*\begin{pmatrix}{\hat{\nu}_{L}}\\{\hat{l}_{L}}\end{pmatrix}
-\kappa\begin{pmatrix}{\hat{\nu}_{L}}\\{\hat{l}_{L}}\end{pmatrix}*\hat{\Lambda}],\\
&\delta_{\hat{\Lambda}}\hat{l}_{R}=ig_{Y}[(-1+\kappa)\hat{\Lambda}*\hat{l}_{R}-\kappa\hat{l}_{R}*\hat{\Lambda}],\\
&\delta_{\hat{\Lambda}}\hat{\Phi}_{d}=ig_{Y}[(-\frac{1}{2}+\kappa)\hat{\Lambda}*\hat{\Phi}_{d}+(1-\kappa)\hat{\Phi}_{d}*\hat{\Lambda}],\\
&\delta_{\hat{\Lambda}}\hat{\Phi}_{s}=i\kappa
g_{Y}\hat{\Lambda}*\hat{\Phi}_{s}-i\kappa
g_{Y}\hat{\Phi}_{s}*\hat{\Lambda}.
\end{split}\end{equation}
Compared with the configuration in Ref. \cite{Calmet1}, not only
the Higgs fields but also the charged lepton fields show  hybrid
feature, where the fields transform under the gauge potentials
from both the left and right sides. Thus, the covariant
derivatives of the lepton fields are given by
\begin{equation}
D_{\mu
L}\hat{\Psi}_{L}=\partial_{\mu}\hat{\Psi}_{L}-ig_{L}\hat{A}_{\mu}^{a}T^{a}*\hat{\Psi}_{L}
-(-\frac{1}{2}+\kappa)g_{Y}\hat{B}_{\mu}*\hat{\Psi}_{L}+i\kappa
g_{Y}\hat{\Psi}_{l}*\hat{B}_{\mu}, \label{c1}\end{equation}
\begin{equation}
D_{\mu R}\hat{l}_{R}=\partial_{\mu}\hat{l}_{R}-i\kappa
g_{Y}B_{\mu}*\hat{l}_{R}+i\kappa g_{Y}\hat{l}_{R}*B_{\mu},
\label{c2}\end{equation} where the $\hat{A}_{\mu}^{a}$ is the
$SU(2)_{L}$ gauge potential and the $g_{L}$ is the coupling
constant. Now we consider the lepton sector of the generalized
mNCSM. Using Eqs. \eqref{c1} and \eqref{c2}, the corresponding
action is
\begin{equation}
\hat{S}_{lepton}=i\int d^4x[\bar{\hat{\Psi}}_{L}\gamma^{\mu}D_{\mu
L}\hat{\Psi}_{L}+\bar{\hat{l}}_{R}\gamma^{\mu}D_{\mu
R}\hat{l}_{R}]. \label{lepton}\end{equation} The next step is to
replace the NC fields in the action with their counterparts in
ordinary space through appropriate Seiberg-Witten mapping.
Usually, the SWM can be derived as perturbative solutions of the
gauge equivalence relation order by order. Recently, the so-called
$\theta$ exact Seiberg-Witten maps involving all orders of the NC
parameter $\theta$ have been obtained by directly solving the
gauge equivalence relation \cite{RH2011, exact2008, exact2011} or
using the recursive formation of the Seiberg-Witten map
\cite{W2011}. Here, we just list the results given in Ref.
\cite{RH2011}:
\begin{equation}\begin{split}
&\hat{\Psi}_{L}=\Psi_{L}-\frac{\theta^{\mu\nu}}{2}(g_{L}A_{\mu}^{a}T^{a}-g_{Y}B_{\mu})\bullet\partial_{\nu}\Psi_{L}
-\theta^{\mu\nu}\kappa
g_{Y}B_{\mu}\star_{2}\partial_{\nu}\Psi_{L}+{\cal O}(a^{2})\Psi_{l},\\
&\hat{l}_{R}=l_{R}+\frac{\theta^{\mu\nu}}{2}g_{Y}B_{\mu}\bullet\partial_{\nu}l_{R}-\theta^{\mu\nu}\kappa
g_{Y}B_{\mu}\star_{2}\partial_{\nu}l_{R}+{\cal O}(A^{2})l_{R},\\
&\hat{\nu}_{R}=\nu_{R}-\theta^{\mu\nu}\kappa
g_{Y}B_{\mu}\star_{2}\partial_{\nu}\nu_{R}+{\cal O}(A^{2})\nu_{R},\\
&\hat{\Phi}_{d}=\Psi_{d}-\frac{\theta^{\mu\nu}}{2}(g_{L}A_{\mu}^{a}T^{a}+g_{Y}B_{\mu})\bullet\partial_{\nu}\Phi_{d}
-\theta^{\mu\nu}(\kappa-1)g_{Y}B_{\mu}\star_{2}\partial_{\nu}\Phi_{d}+{\cal
O}(A^{2})\Phi_{d,}\\
&\hat{\Phi}_{s}=\Psi_{s}-\theta^{\mu\nu}\kappa
g_{Y}B_{\mu}\star_{2}\partial_{\nu}\Phi_{s}+{\cal O}(A^{2})\Phi_{s}
\end{split}\label{SW}\end{equation} with the extended products
$\bullet$ and$\star_{2}$ defined by
\begin{equation}
f\bullet
g:=f\begin{pmatrix}\frac{e^{\frac{i}{2}\theta^{\mu\nu}\overleftarrow{\partial}_{\mu}\overrightarrow{\partial}_{\nu}}-1}
{\frac{i}{2}\theta^{\mu\nu}\overleftarrow{\partial}_{\mu}\overrightarrow{\partial}_{\nu}}\end{pmatrix}g,
\end{equation}
\begin{equation}
f\star_{2}
g:=f\begin{pmatrix}\frac{e^{\frac{i}{2}\theta^{\mu\nu}\overleftarrow{\partial}_{\mu}\overrightarrow{\partial}_{\nu}}-
e^{-\frac{i}{2}\theta^{\mu\nu}\overleftarrow{\partial}_{\mu}\overrightarrow{\partial}_{\nu}}}
{i\theta^{\mu\nu}\overleftarrow{\partial}_{\mu}\overrightarrow{\partial}_{\nu}}\end{pmatrix}g.
\end{equation}
The notation ${\cal O}(A^{2})$ means we only consider the SWM to
the first nontrivial order in the number of gauge potentials.
Theoretically, it is difficult to obtain the analytic expressions
of the higher order terms. We note, however, that for the
processes $e^+e^- \to\mu^+ \mu^-$, the number of gauge fields
taking part in each vertex is not more than one. Thus, we can omit
the higher order contribution. Substituting  Eq. \eqref{SW} into
Eq. \eqref{lepton} and imposing spontaneous symmetry breaking
under the unitary gauge, we can derive all the needed vertices.
The corresponding Feynman rules are
\begin{equation}
-ie\gamma^{\mu}e^{\frac{i}{2}p_{1}\theta p_{2}}-2\kappa
e\gamma^{\mu}\sin(\frac{1}{2}p_{1}\theta p_{2})
\end{equation}
for the photon-electron-electron vertex,  and
\begin{equation}
\frac{ie}{\sin2\theta_{W}}\gamma^{\mu}(C_{V}-C_{A}\gamma^{5})e^{\frac{i}{2}p_{1}\theta
p_{2}}+\frac{2\kappa
e\sin\theta_{W}}{\cos\theta_{W}}\gamma^{\mu}\sin(\frac{1}{2}p_{1}\theta
p_{2})
\end{equation}
for the Z boson-electron-electron vertex. Here $p_{1}$ ($p_{2}$)
is the momentum of the electron ingoing (outgoing) to the vertex;
$p_{1}\theta p_{2}=p_{1}^{\mu}\theta_{\mu\nu}p_{2}^{\nu}$;
$C_{V}=-\frac{1}{2}+2\sin^2\theta_{W}$, $C_{A}=-\frac{1}{2}$, and
$\theta_{W}$ is the Weinberg angle. Since we are only concerned
with the lowest tree-level process, the equations of motion are
applied to the particles in the external lines and the terms
vanish since the on-shell conditions are ignored. The vertex given
above contains not only an exponent-type NC phase factor from the
Moyal product \cite{W2011} but also a periodic term. The new NC
term originates from massive neutrinos, photon-neutrino
interaction, and requirement of gauge invariance. We shall see
that this term leads to phenomenological implications in high
energy processes.

\begin{figure}
 \includegraphics[scale = 0.54]{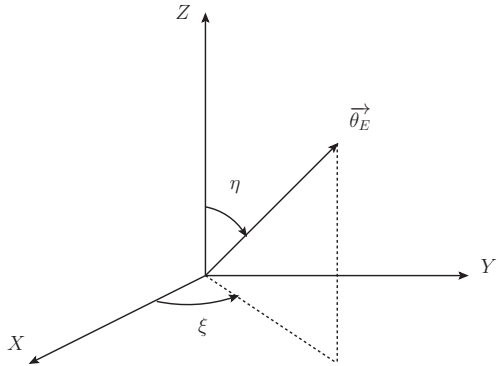}
 \caption{The primary coordinate system ($X-Y-Z$).}
 \label{cor1}\end{figure}

\begin{figure}
 \includegraphics[scale = 0.40]{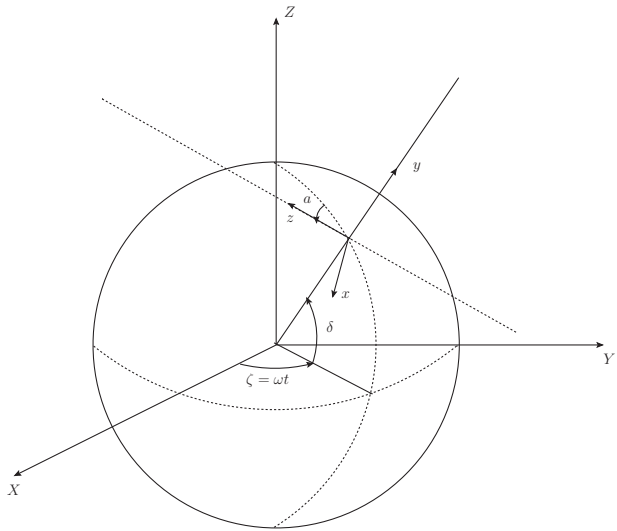}
 \caption{The laboratory coordinate system (x-y-z) on earth.}
 \label{cor2}\end{figure}

\begin{figure}
 \includegraphics[scale = 0.56]{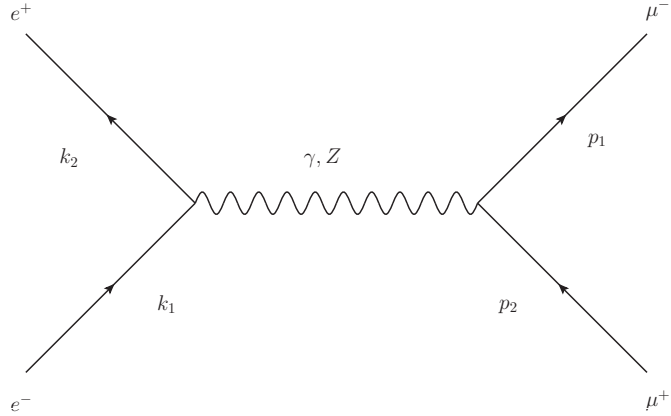}
 \caption{Feynman diagrams for process $e^{+}e^{-}
\to\mu^{+} \mu^{-}$\label{fey}}
 \end{figure}

\section{THE SCATTERING AMPLITUDE IN LABORATORY FRAME}
In this section, we obtain the scattering amplitude of $e^{+}e^{-}
\to\mu^{+} \mu^{-}$ in the laboratory frame. Although carrying the
Lorentz index, the NC parameter $\theta_{\mu\nu}$ is assumed to be
a fundamental constant, which does not change under the Lorentz
transformation but does change with the observer frame. One can
consider both the electric-like vector $\theta_{E}$ and the
magnetic-like vector $\theta_{B}$ to be directionally fixed in a
primary, un-rotational reference. Thus, the earth's motion should
be included when one discusses  phenomenological processes in the
lab frame. Let us define $(\hat{X}, \hat{Y}, \hat{Z})$ to be the
orthonormal basis of this primary system (Fig. \ref{cor1}). In
this frame the NC parameter vector can be written as
\begin{equation}
\theta_{E}=\frac{1}{{\Lambda_{E}^{2}}}(\sin{\eta_{E}}\cos{\xi_{E}}\hat{X}+\sin{\eta_E}\sin{\xi_{E}}\hat{Y}+\cos{\eta_E}\hat{Z}),
\label{theE}\end{equation}
\begin{equation}
\theta_{B}=\frac{1}{{\Lambda_{B}^{2}}}(\sin{\eta_{B}}\cos{\xi_{B}}\hat{X}+\sin{\eta_B}\sin{\xi_{B}}\hat{Y}+\cos{\eta_B}\hat{Z}),
\label{theB}\end{equation} where $\eta$ and $\xi$ denote the NC
polar angular and azimuth angular parameters with $0\leq\eta\leq
\pi$ and $0\leq\xi\leq2\pi $ respectively. However  experiments
are in laboratory frame $(\hat{x},\hat{y},\hat{z})$ on Earth (Fig.
\ref{cor2}). We need a transformation matrix to correlate the two
coordinate systems. Following the notations in Ref.
\cite{Kam2007}, \cite{pkdas}, we have
\begin{eqnarray}
\left(\!\!\!\begin{array}{ccc}
{\hat{X}}\\{\hat{Y}}\\{\hat{Z}}\end{array}\!\!\!\right)=\left(\!\!\!\begin{array}{ccc}
c_{a}s_{\zeta}+s_{\delta}s_{a}c_{\zeta}&c_{\delta}c_{\zeta}&s_{a}s_{\zeta}-s_{\delta}c_{a}c_{\zeta}\\
-c_{a}s_{\zeta}+c_{\delta}s_{a}s_{\zeta}&_{\delta}s_{\zeta}&-s_{a}c_{\zeta}-s_{\delta}c_{a}s_{\zeta}\\
-c_{\delta}s_{a}&s_{\delta}&c_{\delta}c_{a}\end{array}\!\!\!\right)\left(\!\!\!\begin{array}{ccc}
{\hat{x}}\\{\hat{y}}\\{\hat{z}}\end{array}\!\!\!\right),
\label{tran}\end{eqnarray} where the abbreviation $c_{a}=\cos{a}$
etc. is used. The parameters $\delta $ and $a $ denote the
location and orientation
$-\frac{\pi}{2}\leq\delta\leq\frac{\pi}{2}$ and $0\leq a \leq 2
\pi $ of the collider experiment. The parameter $\zeta$ is the
rotation angle defined by $\zeta=\omega t$, where $\omega$ is the
Earth's angular velocity with $\omega=2\pi/{23h56m4.09s}$. Thus,
the collider returns to its original position after a cycle of one
day.

The tree-level Feynman diagram for $e^{+}e^{-} \to\mu^{+} \mu^{-}$
is shown in Fig. \ref{fey}. The process is s-channel and proceed
through photon and Z boson, like in the standard model. Using the
Feynman rules in Sec. 2, the scattering amplitude is
\begin{equation}\begin{split}
M_{\gamma}=&-\frac{ie^{2}}{s}\bar{v}(k_{2})\gamma^{\mu}u(k_{1})\bar{u}(p_{1})\gamma_{\mu}v(p_{2})
[ie^{\frac{i}{2}k_{2}\theta k_{1}}+2\kappa
\sin(\frac{1}{2}k_{2}\theta k_{1})]\\
&\cdot[ie^{\frac{i}{2}p_{1}\theta p_{2}}+2\kappa
\sin(\frac{1}{2}p_{1}\theta p_{2})]
\end{split}\end{equation}
for $\gamma$ mediated interaction and
\begin{equation}\begin{split}
M_{Z}=&\frac{-ie^{2}}{s-m_{Z}^{2}+i\Gamma_{Z}}\bar{v}(k_{2})\gamma^{\mu}[\frac{i}{\sin{2\theta_{W}}}(C_{V}-C_{A}\gamma_{5})e^{\frac{i}{2}k_{2}\theta
k_{1}}+2\kappa \tan_{\theta_{W}}\sin(\frac{1}{2}k_{2}\theta
k_{1})]\\
&\cdot
u(k_{1})\bar{u}(p_{1})\gamma^{\mu}[\frac{i}{\sin{2\theta_{W}}}(C_{V}-C_{A}\gamma_{5})e^{\frac{i}{2}p_{1}\theta
p_{2}}+2\kappa \tan{\theta_{W}}\sin(\frac{1}{2}p_{1}\theta
p_{2})]v(p_{2})
\end{split}\label{momen}\end{equation}
for Z boson mediated interaction, where $k_{1}$, $k_{2}$, $p_{1}$
and $p_{2}$ are the four momenta of the ingoing electron, ingoing
positron, outgoing muon, and outgoing anti-muon, respectively;
$s=(k_{1}+k_{2})^{2}=(p_{1}+p_{2})^{2}$, and $\Gamma_{Z}$ is the
decay width of the $Z$ boson. The total amplitude is
\begin{equation}
M=M_\gamma + M_{Z}.\label{sayid}
\end{equation}
In the center of mass reference,
\begin{equation}\begin{split}
&k_{1}=\begin{pmatrix}\frac{\sqrt{s}}{2},0,0,\frac{\sqrt{s}}{2}\end{pmatrix},\\
&k_{2}=\begin{pmatrix}\frac{\sqrt{s}}{2},0,0,-\frac{\sqrt{s}}{2}\end{pmatrix},\\
&p_{1}=\begin{pmatrix}\frac{\sqrt{s}}{2},\frac{\sqrt{s}}{2}\sin\theta\cos\phi,\frac{\sqrt{s}}{2}\sin\theta\sin\phi,\frac{\sqrt{s}}{2}\cos\theta\end{pmatrix},\\
&p_{2}=\begin{pmatrix}\frac{\sqrt{s}}{2},-\frac{\sqrt{s}}{2}\sin\theta\cos\phi,-\frac{\sqrt{s}}{2}\sin\theta\sin\phi,-\frac{\sqrt{s}}{2}\cos\theta\end{pmatrix},\\
\end{split}\end{equation}
where $\theta$ is the polar angle and $\phi$ is the azimuthal
angle with respect to the initial beam direction along the z-axis.
Using Eqs. \eqref{theE}, \eqref{theB}, \eqref{tran} and
\eqref{momen}, we obtain
\begin{equation}\begin{split}
&k_{2}\theta k_{1}=-\frac{s}{2\Lambda^{2}_{NC}}\Theta_{E}^{z},\\
&p_{2}\theta p_{1}=-\frac{s}{2\Lambda^{2}_{NC}}
(\sin\theta\cos\phi\Theta_{E}^{x}+\sin\theta\sin\phi\Theta_{E}^{y}+\cos\theta\Theta_{E}^{z})
\end{split}\end{equation}
with
\begin{equation}\begin{split}
&\Theta_{E}^{x}=s_{\eta}c_{\xi}(c_{a}s_{\zeta}+s_{\delta}s_{a}c_{\zeta})+
s_{\eta}s_{\xi}(-c_{a}s_{\zeta}+c_{\delta}s_{a}s_{\zeta})-c_{\eta}c_{\delta}c_{a},\\
&\Theta_{E}^{y}=s_{\eta}c_{\xi}c_{\delta}c_{\zeta}+s_{\eta}s_{\xi}c_{\delta}c_{\zeta}+c_{\eta}c_{\delta},\\
&\Theta_{E}^{z}=s_{\eta}c_{\xi}(s_{a}s_{\zeta}-s_{\delta}c_{a}c_{\zeta})+s_{\eta}s_{\xi}(-s_{a}c_{\zeta}-s_{\delta}c_{a}s_{\zeta})
+c_{\eta}c_{\delta}c_{a}.
\end{split}\end{equation}
One can see that the process $e^+e^- \to\mu^+ \mu^-$ is only
sensitive to  $\theta_{E}$. Then, the squared amplitude under
spin-averaging is
\begin{equation}
\overline{|M|^{2}}=
\overline{|M_{\gamma}|^{2}}+\overline{|M_{Z}|^{2}}+2\overline{Re(M_{\gamma}M^{\dagger}_{Z})}.
\end{equation}
Using the tracing technique, the elements of squared amplitude are
\begin{equation}
\overline{|M_{\gamma}|^{2}}=\frac{e^{4}}{4s^{2}}[(k_{2}\cdot
p_{1})(k_{1}\cdot p_{2})+(k_{2}\cdot p_{2})(k_{1}\cdot p_{1})]AB,
\end{equation}
\begin{equation}
\overline{|M_{Z}|^{2}}=\frac{e^{4}}{4[(s-m_{Z}^{2})^{2}+\Gamma_{Z}^{2}m_{Z}^{2}]}[C_{+}(k_{2}\cdot
p_{1})(k_{1}\cdot p_{2})+C_{-}(k_{2}\cdot p_{2})(k_{1}\cdot
p_{1})],
\end{equation}
\begin{equation}
2\overline{Re(M_{\gamma}M^{\dagger}_{Z})}=\frac{e^{4}(s-m_{Z}^{2})}{2s[(s-m_{Z}^{2})^{2}+\Gamma_{Z}^{2}m_{Z}^{2}]}[D_{+}(k_{2}\cdot
p_{1})(k_{1}\cdot p_{2})+D_{-}(k_{2}\cdot p_{2})(k_{1}\cdot
p_{1})],
\end{equation}
where
\begin{equation}
A=32+128\kappa(\kappa-1)\sin^{2}(\frac{1}{2}k_{1}\theta k_{2}),
\end{equation}
\begin{equation}
B=32+128\kappa(\kappa-1)\sin^{2}(\frac{1}{2}p_{2}\theta k_{1}),
\end{equation}
\begin{equation}\begin{split}
C_{\pm}&=\frac{1}{\sin^{4}2\theta_{W}}\begin{bmatrix}32(C_{V}^{2}+C_{A}^{2})^{2}\pm128C_{V}^{2}C_{A}^{2}\end{bmatrix}\\
&+\begin{bmatrix}-\frac{128\kappa
C_{V}\tan\theta_{W}}{\sin^{3}2\theta_{W}}(C_{V}^{2}+C_{A}^{2}\pm
2C_{A}^{2})
+\frac{128\kappa^{2}\tan^{2}\theta_{W}}{\sin^{2}2\theta_{W}}\end{bmatrix}\sin^{2}(\frac{1}{2}k_{1}\theta
k_{2})\\
&+\begin{bmatrix}-\frac{128\kappa
C_{V}\tan\theta_{W}}{\sin^{3}2\theta_{W}}(C_{V}^{2}+C_{A}^{2}\pm
2C_{A}^{2})
+\frac{128\kappa^{2}\tan^{2}\theta_{W}}{\sin^{2}2\theta_{W}}(C_{V}^{2}+C_{A}^{2})\end{bmatrix}\sin^{2}(\frac{1}{2}p_{2}\theta
p_{1})\\
&+\begin{bmatrix}\frac{512\kappa^{2}\tan^{2}\theta_{W}}{\sin^{2}2\theta_{W}}(C_{V}^{2}\pm
C_{A}^{2})
-\frac{1024\kappa^{3}C_{V}\tan^{3}\theta_{W}}{\sin2\theta_{W}}+512\kappa^{4}\tan^{4}\theta_{W}\end{bmatrix}
\sin^{2}(\frac{1}{2}k_{1}\theta
k_{2})\\
&\cdot\sin^{2}(\frac{1}{2}p_{2}\theta p_{1}),
\end{split}\end{equation}
\begin{equation}\begin{split}
D_{\pm}=&\frac{32}{\sin^{2}2\theta_{W}}(C_{V}^{2}\pm
C_{A}^{2})+\begin{bmatrix}\frac{128\kappa(\kappa-0.5)
C_{V}\tan\theta_{W}}{\sin2\theta_{W}}-\frac{64\kappa(C_{V}^{2}\pm
2C_{A}^{2})
}{\sin^{2}2\theta_{W}}\end{bmatrix}\sin^{2}(\frac{1}{2}k_{1}\theta
k_{2})\\
&+\begin{bmatrix}\frac{128\kappa(\kappa-0.5)
C_{V}\tan\theta_{W}}{\sin2\theta_{W}}-\frac{64\kappa(C_{V}^{2}\pm
2C_{A}^{2})
}{\sin^{2}2\theta_{W}}\end{bmatrix}\sin^{2}(\frac{1}{2}p_{2}\theta
p_{1})\\
&+\begin{bmatrix}\frac{256\kappa^{2}
C_{V}\tan\theta_{W}}{\sin2\theta_{W}}-128\kappa^{2}\tan^{2}\theta_{W}\begin{pmatrix}1
+\frac{C_{V}^{2}\pm
C_{A}^{2}}{\sin^{2}2\theta_{W}}\end{pmatrix}\end{bmatrix}
\cos(\frac{1}{2}k_{1}\theta k_{2})\cos(\frac{1}{2}p_{2}\theta
p_{1})\\& \sin(\frac{1}{2}k_{1}\theta
k_{2})\sin(\frac{1}{2}p_{2}\theta p_{1})\\
&+\begin{bmatrix}\frac{256\kappa^{2}(1-2\kappa)
C_{V}\tan\theta_{W}}{\sin2\theta_{W}}+128\kappa^{2}\tan^{2}\theta_{W}\begin{pmatrix}1
+\frac{C_{V}^{2}\pm
C_{A}^{2}}{\sin^{2}2\theta_{W}}\end{pmatrix}+512\kappa^{3}(\kappa-1)\tan^{2}\theta_{W}\end{bmatrix}\\
& \sin^{2}(\frac{1}{2}k_{1}\theta
k_{2})\sin^{2}(\frac{1}{2}p_{2}\theta p_{1}).
\end{split}\end{equation}
In the calculation, the FeynCalc package of Mathematica
\cite{COMPU} is used and  the fermion mass is neglected in the
high energy limit.

\begin{figure}
 \includegraphics[scale = 0.56]{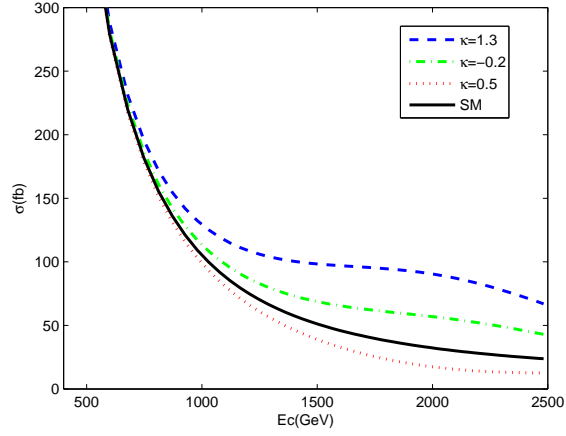}%
\caption {The total cross section $\langle\sigma\rangle_{T}$ after
time averaging
 as a function of $E_c$ in ordinary
space-time and noncommutative space-time for $\Lambda_{NC}=800$
GeV, $\eta=\frac{\pi}{4}$, and $\kappa=$1.3, -0.2, and 0.5. }
\label{ttp1}
 \end{figure}

\begin{figure}
 \includegraphics[scale = 0.56]{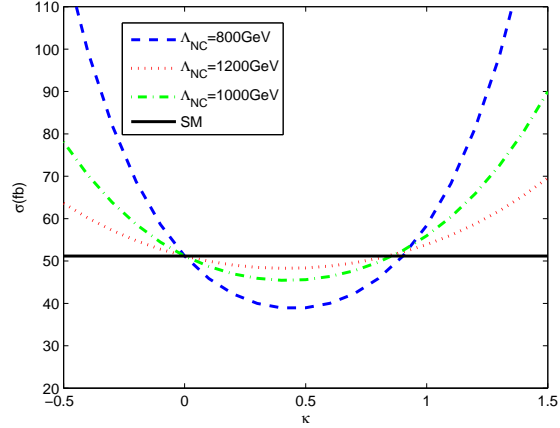}%
\caption {The total cross section $\langle\sigma\rangle_{T}$ after
time averaging
 as a function of $\kappa$ for
$E_{c}=1500$ GeV, $\eta=\frac{\pi}{4}$ rad, and $\Lambda_{NC}=800,
1000, \mathrm{and } 1200$ GeV. } \label{ttp2}
 \end{figure}

\begin{figure}
 \includegraphics[scale = 0.56]{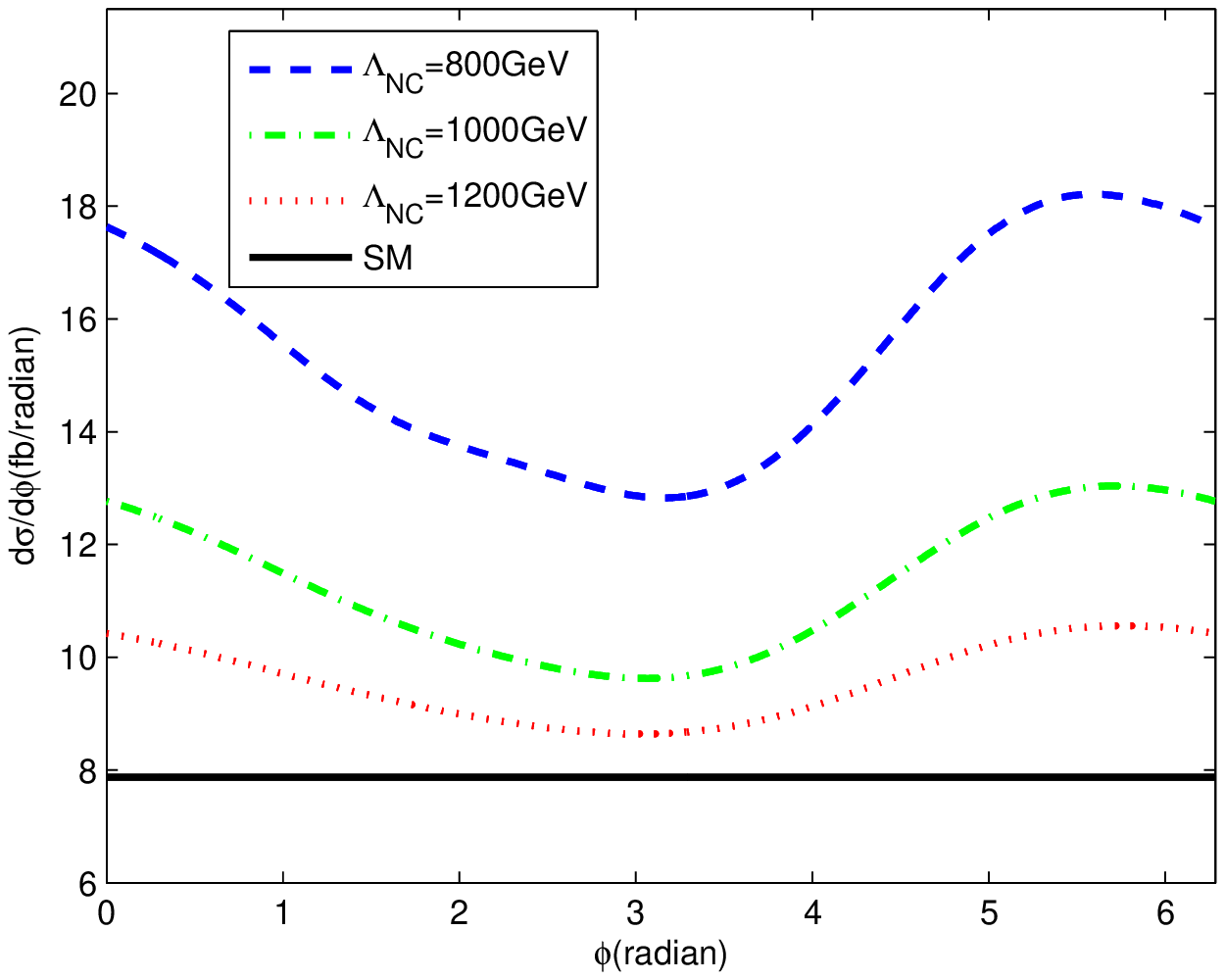}%
\caption {The time averaged
$\langle\frac{d\sigma}{d\phi}\rangle_{T}$ as a function of $\phi$
for $E_{c}=1500$ GeV, $\kappa$=1.3, $\eta=\frac{\pi}{4}$ rad, and
$\Lambda_{NC}=800, 1000, \mathrm{and } 1200$ GeV.} \label{ttp3}
 \end{figure}

\begin{figure}
 \includegraphics[scale = 0.56]{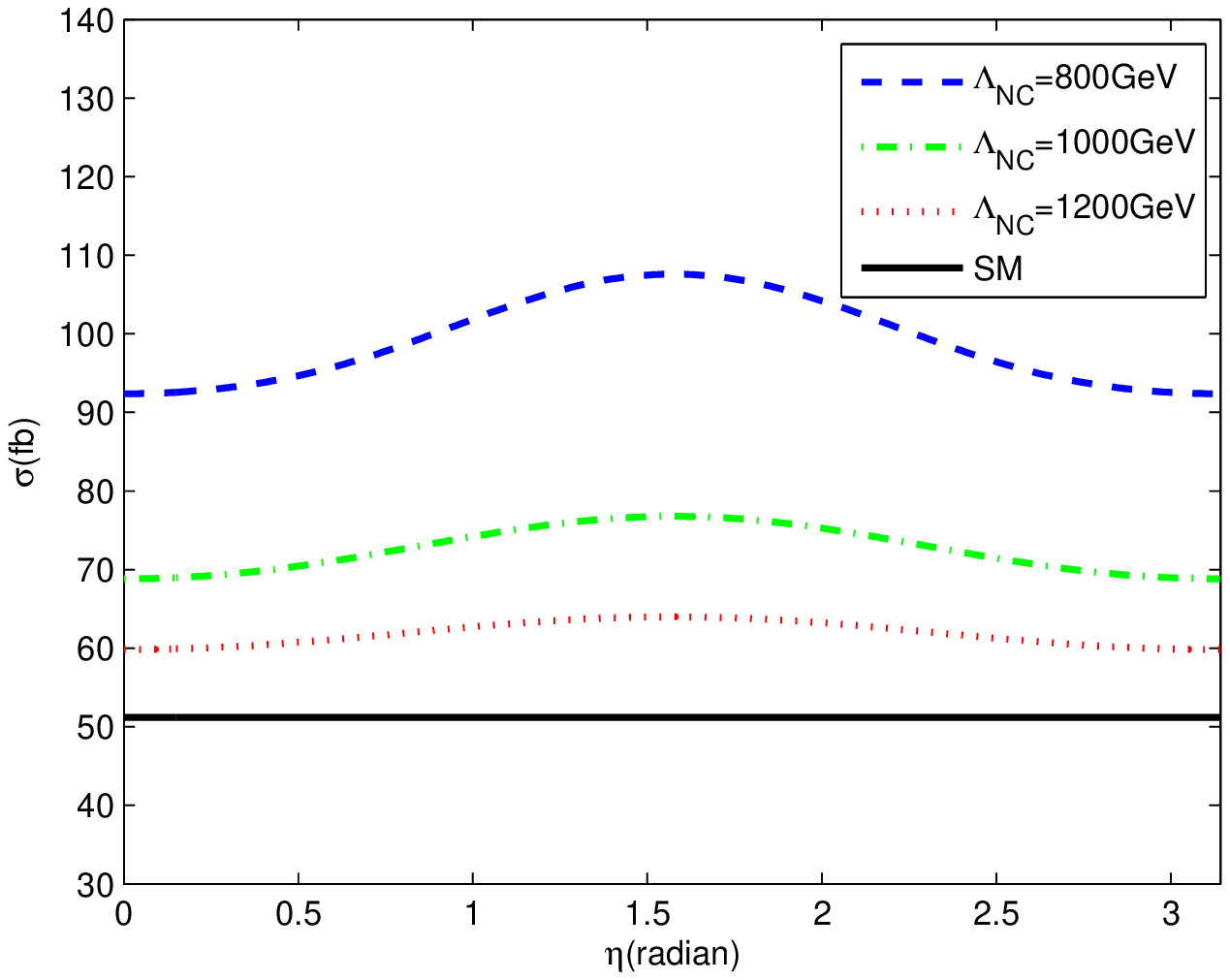}%
\caption {The time averaged $\langle\sigma\rangle_{T}$ as a
function of $\eta$ for $E_{c}=1500$ GeV, $\kappa$=1.3,
$\eta=\frac{\pi}{4}$ rad, and $\Lambda_{NC}=800, 1000, \mathrm{and
} 1200$ GeV. } \label{ttp4}
 \end{figure}

\begin{figure}
 \includegraphics[scale = 0.56]{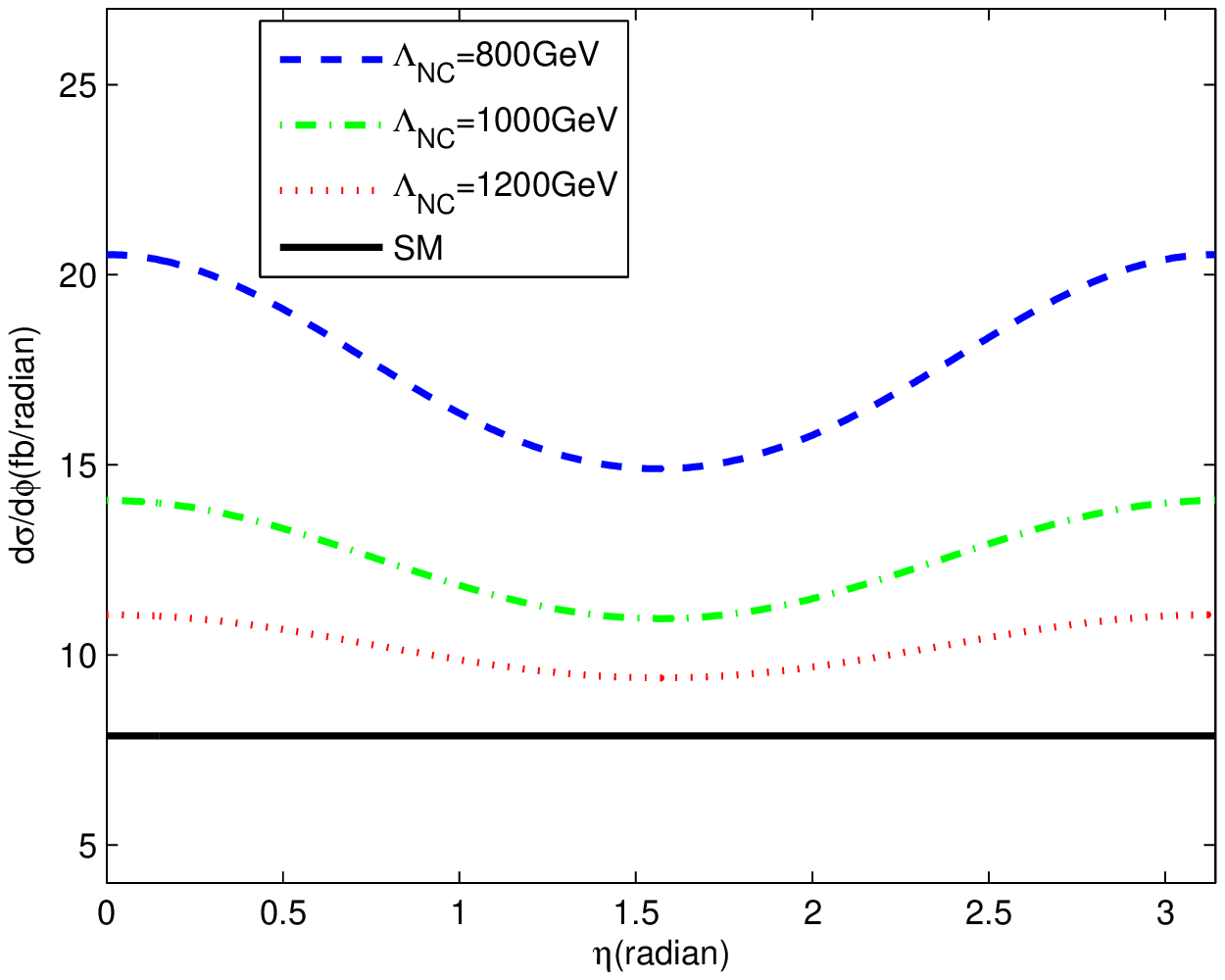}%
\caption {The time averaged
$\langle\frac{d\sigma}{d\phi}\rangle_{T}$ as a function of $\eta$
for $E_{c}=1500$ GeV, $\kappa$=1.3, $\phi=5$ rad, and
$\Lambda_{NC}=800, 1000, \mathrm{and } 1200$ GeV. } \label{ttp5}
 \end{figure}

%---------------------------------------------------------------------------------------

\section{NUMERICAL ANALYSIS}
In this section, we analyze the total cross section and angular
distribution of the process  $e^{+}e^{-} \to\mu^{+} \mu^{-}$ in
the framework of the generalized NCSM. Because of the Earth's
rotation, it is difficult to get  time-dependent data from the
collider, so that the observable here should be averaged over a
full day in order to compare them with the experimental results.
The time-averaged differential cross section is
\begin{equation}
\langle\frac{d\sigma}{d\phi}\rangle_{T}=
\frac{1}{T_{day}}\int_{0}^{T_{day}}\frac{d\sigma}{d\cos\theta
d\phi}dt,
\end{equation}
where the differential cross section for the two body process is
given by
\begin{equation}
\frac{d\sigma}{d\cos\theta
d\phi}=\frac{1}{64\pi^{2}s}\overline{|M|}^{2}.
\end{equation}
After integrating, we get the timed-averaged total cross section
\begin{equation}
\langle\sigma\rangle_{T}=\frac{1}{T_{day}}\int_{0}^{T_{day}}dt\int_{-1}^{1}d(\cos\theta)\int_{0}^{2\pi}d\phi
\frac{d\sigma}{d\cos\theta d\phi}.
\end{equation}

\subsection{Time averaged total cross section and angular distribution}
In Fig. \ref{ttp1}, we show the ordinary total cross section
$\sigma_{0}$ and the NC corrected total cross section
$\langle\sigma\rangle_{T}$ as function of
 the collision energy $E_c(=\sqrt{s})$ for
 $\Lambda_{NC}=800$ GeV and $\kappa=1.3, 0.5, \mathrm{and} -0.2$.
 The solid curve corresponds to the SM case.
 In our numerical analysis,
 we set the location coordinate of the laboratory frame at $(\delta, a)=(\frac{\pi}{4},
 \frac{\pi}{4})$, which is the location of the OPAL experiment at LEP.
One can find from the figure that the NC effect causes significant
deviation to the total cross section when $E_c$ is high enough.
Interestingly, here $\langle\sigma\rangle_{T}$ is sensitive to
both $\Lambda_{NC}$ and $\kappa$. When the NC scale parameter
$\Lambda_{NC}$ is fixed, the total cross section can be enhanced
or suppressed for different values of $\kappa$. To see more about
this, we present $\langle\sigma\rangle_{T}$ as a function of the
parameter $\kappa$ in Fig. \ref{ttp2} where $E_c(=\sqrt{s})$ is
fixed at 1500 GeV, for $\Lambda_{NC}=800$ GeV, 1000 GeV and 1200
GeV. The horizontal line corresponds to the SM case. For
simplicity, here we assume that  $\kappa$ varies between -0.5 to
1.5. The figure shows that for a fixed collision energy the cross
section has a parabolic dependence on $\kappa$. Despite different
NC scale parameters at around 1 TeV, the total cross section is
greatly enhanced when $\kappa$ is  at [-0.5, 0] and about [0.9,
1.5]. If $\kappa$ is at about [0, 0.9], the cross section will be
suppressed. We note that a similar picture also appears in Ref.
\cite{W2011}, in which the deformation of covariant derivatives is
due to\textsl{a priori} assumption and only limited to the Higgs
sector. Thus, in that case no NC effect is manifested in the
process $e^+e^- \to\mu^+ \mu^-$. In Sec. 2, we have shown that the
Feynman rules of electron-photon interaction and electron- Z boson
interaction contain sin-type deformation coming from the
consistency between the neutrino-photon interaction and seesaw
extension of SM. As one of our main results, we show that this
deformation indeed predicts interesting deviation from the total
cross section.

We plot the the azimuthal angular distribution
$\frac{d\sigma}{d\phi}$ in Fig.\ref{ttp3} for $\Lambda_{NC}=800$
GeV, 1000 GeV and 1200 GeV. The horizontal line is for the SM
case. Here the collision energy $E_c$ is $1500$ GeV and
$\eta=\frac{\pi}{4}$. One can see from the figure that
$\frac{d\sigma}{d\phi}$ is anisotropic. This is because the
space-time noncommutativity is spontaneous Lorentz violation and
breaking of rotational invariance. In our analysis, all three
curves reach their maxima at around $\phi=5.58$ rad and their
minima at $\phi=3.18$ rad. This unique feature can help us to
identify the NC effect from the other effects.

\subsection{Time averaged total cross section and angular distribution as a function of $\eta$}
Since $\theta_{E}$ is assumed to be a stationary vector fixed in
the primary frame, any physical value calculated in NC space-time
is not only sensitive to $\Lambda_{NC}$ but also to its direction
parameter $(\eta, \xi)$. After taking the average over a full day
rotation, $\eta$ remains. In Fig. \ref{ttp4}, we present the
$\langle\sigma\rangle_{T}$ and $\frac{d\sigma}{d\phi}$ as
functions of $\eta$, for $E_c(=\sqrt{s})=$ 1500 GeV and $\kappa=$
1.3. The curves show a positive kurtosis distribution for the
whole range of $\eta$, and the horizonal solid line corresponds to
the SM case. The maximum NC correction of all curves appear at
$\eta=1.53$ rad. Thus, one can detect the NC effect for any
$\theta_{E}$.

In Fig. \ref{ttp5}, we plot
$\langle\frac{d\sigma}{d\phi}\rangle_{T}$ as  function of $\eta$
for $\Lambda_{NC}=800, 1000, 1200$ GeV. Here we fix the $\phi$ at
5 rad, $\kappa=1.3$ and collision energy $E_{c}=$ 1500 GeV.
Different from $\langle\sigma\rangle_{T}$,  the minima of the NC
corrections are round $\phi=1.53$ rad.

\section{conclusion and discussion}
In this work, we have considered a generalized noncommutative
standard model, in which the massive neutrino and direct
neutrino-photon interaction are included. It is found that the
direct neutrino-photon interaction in NC space-time will have
effect in the lepton sector and introduce hybrid gauge
transformation by  requiring gauge invariance. As an application,
we study the TeV phenomenology of $e^+e^- \to\mu^+ \mu^-$
scattering at $e^+e^-$ linear colliding. In NCSM without hybrid
gauge transformation, it was found that when all  orders of
$\theta$ are included, there is no NC correction to the squared
amplitude of $e^+e^- \to\mu^+ \mu^-$ process. In the generalized
NCSM, however, after deriving the corresponding Feynman rules, we
find that there are additional sin-type deformations compared to
the ones given in Ref. \cite{W2011}. These deformations lead to
the nontrivial phenomenological implication that the cross section
of $e^+e^- \to\mu^+ \mu^-$ process  can also have the NC effect,
which is potentially detectable in the future International Linear
Collider (ILC). The Earth's rotation is also included. The cross
section and angular distribution are analyzed in the laboratory
frame. Pair production of muons via $e^{+}e^{-}$ collision in the
ILC should provide an ideal opportunity for probing not only the
NC space-time, but also the mathematical structure of the
corresponding gauge theory.

Whether the deformed terms exist and how can one fix the value of
$\kappa$ is still an open question. This may be because we still
have not enough information on the renormalizability of the NC
quantum field theory, where  freedom of the deformation terms can
be used to cancel the UV divergence \cite{Buric2002}. It is
expected that further work on the renormalizability can remove
these ambiguity. Before that one can treat it as an effective
theory and the phenomenological study can give constraints on it,
as has been done in the study of quarkonia decays
\cite{Tamarit09}.

It is feasible to investigate other standard scatterings such as
the Moller and Bhabha scattering. Although these processes are
more kinematically complicated, they are ideal cases for detecting
noncommutativity between space and space. This topic is
interesting and deserve further study.

\begin{acknowledgments}
Weijian Wang would like to thank Jie Yin for helpful discussions.
This work is supported in part by the funds from NSFC under Grant
No.11075140 and the Fundamental Research Funds for the Central
University.
\end{acknowledgments}

\end{document}